\documentclass[12pt]{article}
\usepackage[dvips]{graphicx}
\usepackage{amssymb}
\usepackage{amsmath}
\usepackage{epsfig}
\usepackage{cite}
\usepackage{hyperref}

\numberwithin{equation}{section}
\numberwithin{table}{section}

\def\beq{\begin{equation}}
\def\eeq{\end{equation}}
\def\be{\begin{equation}}
\def\ee{\end{equation}}
\def\bea{\begin{eqnarray}}
\def\eea{\end{eqnarray}}

\DeclareMathOperator{\rk}{rk}
\def\SE{\text{SE}}

\def\del{\partial}

\def\Re{{\rm Re\,}}
\def\Im{{\rm Im\,}}

\def\g{{\mathfrak{g}}}

\def\vev#1{{\langle #1\rangle}}

\DeclareRobustCommand{\SkipTocEntry}[4]{}
\RequirePackage{color}

\def\vev#1{{\langle #1\rangle}}

\newcommand{\cO}{\mathcal{O}}
\newcommand{\cT}{\mathcal{T}}
\newcommand{\cE}{\mathcal{E}}

\newcommand{\cD}{\mathcal{D}}
\newcommand{\cL}{\mathcal{L}}

\newcommand{\cM}{\mathcal{M}}
\newcommand{\cN}{\mathcal{N}}

\newcommand{\cG}{\mathcal{G}}
\newcommand{\cA}{\mathcal{A}}
\newcommand{\cH}{\mathcal{H}}
\newcommand{\cB}{\mathcal{B}}

\newcommand{\bbZ}{\mathbb{Z}}
\newcommand{\bbR}{\mathbb{R}}
\newcommand{\bbC}{\mathbb{C}}

\def\AdS{\textrm{AdS}}
\def\dim{\textrm{dim }}

\begin{document}
\begin{titlepage}
\begin{center}
\rightline{\small ZMP-HH/16-25}

\vskip 1cm

{\Large \bf $\AdS_{5}$ vacua from type IIB supergravity on $T^{1,1}$}
\vskip 1.2cm

{\bf  Jan Louis 
and Constantin Muranaka}

\vskip 0.8cm

{\em Fachbereich Physik der Universit\"at Hamburg, Luruper Chaussee 149, 22761 Hamburg, Germany}
\vskip 0.2cm

and 

\vskip 0.2cm

{\em Zentrum f\"ur Mathematische Physik,
Universit\"at Hamburg,\\
Bundesstrasse 55, D-20146 Hamburg, Germany}
\vskip 0.3cm

\vskip 0.3cm

{\tt jan.louis@desy.de, constantin.muranaka@desy.de}

\end{center}

\vskip 1cm

\begin{center} {\bf ABSTRACT } \end{center}

We study maximally supersymmetric Anti-de Sitter backgrounds in consistent $\cN=2$ truncations of type IIB supergravity compactified on the Sasaki-Einstein manifold $T^{1,1}$. In particular, we focus on truncations that contain fields coming from the nontrivial second and third cohomology forms on $T^{1,1}$. These give rise to $\cN=2$ supergravity coupled to two vector- and two hypermultiplets (Betti-vector truncation) or one vector- and three hypermultiplets (Betti-hyper truncation), respectively. We find that both truncations admit $\AdS_{5}$ backgrounds with the gauge group always being broken but containing at least an $U(1)_{R}$ factor. Moreover, in both cases we show that the moduli space of AdS vacua is nontrivial and of maximal dimension. Finally, we explicitly compute the metrics on these moduli spaces.

\noindent

\bigskip

\vfill

November 2016

\end{titlepage}




\section{Introduction}


Supersymmetric solutions of type IIB supergravity with AdS factors preserving 8 real supercharges have been extensively studied in the past \cite{Gauntlett:2005ww,Gabella:2009wu,Coimbra:2015nha,Ashmore:2016oug,Ashmore:2016qvs,Grana:2016dyl}. They are related to gauge theories on the boundary of AdS by the AdS/CFT correspondence \footnote{See \cite{Maldacena:1997re,Morrison:1998cs} for earlier work and \cite{Polchinski:2010hw} for a recent review.}. On the other hand, references \cite{Tachikawa:2005tq, Louis:2016qca,Louis:2015dca,Louis:2016tnz,Louis:2015mka,Louis:2014gxa,deAlwis:2013jaa} studied AdS vacua in different dimensions and with varying amounts of supersymmetry purely from a supergravity perspective, without any relation to compactifications of higher-dimensional theories. It was found explicitly in every case that these AdS backgrounds have properties which agree with the associated dual gauge theories. For example, all studied backgrounds admit an R-symmetry in the background that precisely matches the gauge theory R-symmetry. Moreover, it was explicitly shown in each case that moduli spaces of AdS backgrounds satisfy the same conditions as the conformal manifolds of the dual gauge theories.

In this work we study five-dimensional $\cN=2$ AdS backgrounds obtained via consistent truncations from ten-dimensional compactifications of type IIB supergravity. In particular, we will focus on the well-known background of the form $AdS_{5}\times T^{1,1}$, where $T^{1,1}=\frac{SU(2)\times SU(2)}{U(1)}$ is the Sasaki-Einstein manifold underlying the conifold. Due to the coset structure of $T^{1,1}$, one can apply an argument of \cite{Duff:1985jd}: retaining only singlets of a transitively acting isometry subgroup in a Kaluza-Klein reduction is a necessary condition for the consistency of the truncation. Since $T^{1,1}$ is a coset space, the group $SU(2)\times SU(2)$ acts transitively and is a subgroup of the isometry. It was shown in \cite{Cassani:2010na,Bena:2010pr} that truncating to singlets of this group indeed is consistent and leads to a five-dimensional gauged $\cN=4$ supergravity theory coupled to three vector multiplets. In particular, to study $\cN=2$ AdS backgrounds we will be interested in certain $\cN=2$ subtruncations of this $\cN=4$ theory developed in \cite{Cassani:2010na,Halmagyi:2011yd}.

Since the $AdS_{5}\times T^{1,1}$ background is dual to the Klebanov-Witten theory \cite{Klebanov:1998hh}, this solution has been extensively studied in the past. For example, the moduli space of the full ten-dimensional theory is known to be complex five-dimensional \cite{Klebanov:1998hh,Benvenuti:2005wi,Ardehali:2014zfa,Ashmore:2016qvs,Ashmore:2016oug} and the moduli have been identified. There is always one modulus corresponding to the Axion-Dilaton $\tau$ and another coming from the vacuum expectation value (VEV) of the complex B-field of type IIB supergravity integrated over the nontrivial two-cycle of $T^{1,1}$.\footnote{This comes from the fact that $T^{1,1}$ is diffeomorphic to $S^{2}\times S^{3}$ and thus has non-vanishing second Betti-number \cite{Cassani:2010na,Bena:2010pr}.} Moreover the remaining three complex moduli (one of which is the deformation identified in \cite{Lunin:2005jy}) transform as a triplet under the $SU(2)\times SU(2)$ in the isometry group. However, it has been shown in \cite{Hoxha:2000jf} that it is inconsistent to keep non-singlet representations of $SU(2)\times SU(2)$ in a truncation and thus only the first two moduli are expected to show up in the truncated theories.

The aim of this paper is to find the conditions on AdS backgrounds and their moduli spaces in the consistent $\cN=2$ truncations of $T^{1,1}$ supergravity. To this end, we focus on the truncations known as the Betti-vector truncation\footnote{This truncations was recently studied in the context of holographic duals of c-extremization \cite{Amariti:2016mnz}.} and the Betti-hyper truncation in \cite{Halmagyi:2011yd}, since they contain multiplets associated with the topology of $T^{1,1}\cong S^{2}\times S^{3}$. The Betti-vector truncation contains gravity coupled to two vector multiplets and two hypermultiplets, while the Betti-hyper truncation contains gravity coupled to one vector multiplet and three hypermultiplets. We then apply the methods developed in \cite{Louis:2016qca} to these truncations. We find that both truncations admit AdS vacua with an unbroken $U(1)_{R}$ symmetry in the background. Moreover, we explicitly compute the metric on the moduli spaces. In the case of the Betti-vector truncation, we find that the moduli space $\cM^{BV}$ is spanned by the Axion-Dilaton $\tau$ and $\cM^{BV}=\cH$ is the upper half-plane $\cH$. For the Betti-hyper truncation we compute a complex two-dimensional moduli space $\cM^{BH}$ that is given by a torus bundle with base space parametrized again by $\tau$. In particular, this reproduces the result of \cite{Klebanov:1998hh} that the moduli in question are the Axion-Dilaton and a complex scalar that parametrizes a torus. However, the metric on the moduli space is not a direct product but a nontrivial fibration known as the universal elliptic curve $\cE= (\bbC\times \cH)/\bbZ^{2}$ \cite{RHain}. 

The rest of this paper is organized as follows: in section \ref{sugra} we briefly review gauged $\cN=2$ supergravity in five dimensions and the conditions on $\AdS_{5}$ vacua in these theories. In section \ref{truncations} we introduce the relevant $\cN=2$ truncations on $T^{1,1}$ and collect the relevant data. We then use this to compute the conditions on AdS vacua and moduli spaces in these truncations. Finally, we conclude and discuss our results in section \ref{conclusion}.


\section{$\AdS_{5}$ vacua in $\cN=2$ gauged supergravity}\label{sugra}

Let us start by reviewing $\AdS_{5}$ vacua of gauged $\cN=2$, $d=5$ supergravity as discussed in \cite{Tachikawa:2005tq, Louis:2016qca}. To this end we introduce gauged $\cN=2$ supergravity as discussed in \cite{Gunaydin:2000xk,Bergshoeff:2002qk,Bergshoeff:2004kh}. In this paper we will consider theories with the following field content: the gravity multiplet
\begin{equation}
 \{g_{\mu\nu}, \psi_{\mu}^{\cA}, A_{\mu}^{0}\}\ , \quad \mu,\nu = 0,...,4\ , \quad \cA=1,2\ ,
\end{equation}
containing the spacetime metric $g_{\mu\nu}$, an $SU(2)_{R}$ doublet of symplectic Majorana fermions $\psi_{\mu}^{\cA}$ and a vector $A_{\mu}^{0}$ called the graviphoton. Additionally, we allow for $n_{V}$ vector multiplets $\{A_{\mu}, \lambda^{\cA}, \phi\}$  that transform in the adjoint of the gauge group, containing a vector $A_{\mu}$, a doublet of gauginos $\lambda^{\cA}$ and a real scalar $\phi$. Including the graviphoton, we label the vector fields by $I,J=0,...,n_{V}$ and their associated scalars by $i,j = 1,...,n_{v}$.\footnote{In \cite{Louis:2016qca} we also considered tensor multiplets which arise by dualising vector fields that transform in a different representation of the gauge group. Since these do not appear in the truncations that we will analyze later, we do not include them here.} Note that there is no scalar associated to the graviphoton. Moreover, we include $n_{H}$ hypermultiplets 
\begin{equation}
 \{q^{u}, \zeta^{\alpha}\}\ , \quad u=1,2,...,4n_{H}\ , \quad \alpha=1,2,...,2n_{H}\ ,
\end{equation}
containing $4n_{H}$ real scalars $q^{u}$ and $2n_{H}$ hyperini $\zeta^{\alpha}$.

The scalar fields can be interpreted as coordinate charts from spacetime $M_{5}$ to a target space $\cT$, $\phi^{i}\otimes q^{u}:M_{5}\longrightarrow \cT$. Locally $\cT \cong \cT_{V} \times \cT_{H}$ where the first factor is a very special real manifold of dimensions $n_{V}$. This is defined as a hypersurface in an $(n_{V}+1)$-dimensional real manifold $\cH$ by
\begin{equation}
 P(h^{I}(\phi))=C_{IJK}h^{I}h^{J}h^{K}=1\ ,
\end{equation}
where $h^{I}$ are the coordinates on $\cH$ and $P(h^{I}(\phi))$ is a cubic homogeneous polynomial with $C_{IJK}$ constant and completely symmetric. The index $I$ can be lowered via
\begin{equation}\label{hindex}
 h_{I}=C_{IJK}h^{J}h^{K}\ .
\end{equation}

The second factor of $\cT$ is a quaternionic K\"ahler manifold of real dimension $4n_{H}$ (see \cite{Andrianopoli:1996cm} for a more detailed introduction). This is equipped with a metric $G_{uv}$ and a triplet of almost complex structures $\vec{J}$ such that $J^{1}J^{2}=J^{3}$. In particular, $G_{uv}$ is hermitian with respect to all of these almost complex structures and one can define the associated two-forms $\vec\omega_{uv}:=G_{uw}\vec{J}^{w}_{v}$. In what follows we need to know more about the action of the gauge group $G$ on the target space of the hypermultiplet scalars. This is given by Killing vectors $k_{I}^{u}$ on $\cT_{H}$ that respect the quaternionic structure, i.e.\ that are triholomorphic. This implies the existence of a triplet of moment maps $\vec{\mu}_{I}$ which satisfy
\begin{equation}
 \tfrac{1}{2}\vec{\omega}_{uv}k_{I}^{v}=-\nabla_{u}\vec{\mu}_{I}\ ,
\end{equation}
where $\nabla_u$ is a connection that combines the Levi-Civita connection for $G_{uv}$ and an $SU(2)$-connection. Moreover the covariant derivatives of the scalars $q^{u}$ are given by
\begin{equation}\label{derivatives}
\cD_{\mu}q^{u}=\del_{\mu}q^{u}+k_{I}^{u}A^{I}_{\mu}\ .
\end{equation}

Equipped with the above data one can then define the scalar potential $V(\phi^{i}, q^{u})$ and the supersymmetry variations of the fermions. In a maximally supersymmetric AdS background the latter have to vanish, i.e.\
\begin{equation}\label{Fermions}
 \vev{\delta\psi_{\mu}^{\cA}}=\vev{\delta\lambda^{\cA}}=\vev{\delta\zeta^{\alpha}}=0\ .
\end{equation}
Then the scalar potential is nonzero in the background, $\vev{V}=\Lambda$ where $\Lambda \in \bbR$ is the cosmological constant. It has been shown in \cite{Louis:2016qca} that the conditions \eqref{Fermions} can be expressed in terms of the moment maps and Killing vectors on the scalar manifold,
\begin{equation}\label{background}
 \vev{h^{I}k_{I}}=0\ , \quad \vev{\vec{\mu}_{I}} = \lambda \vev{h_{I}} \vec{v}\ ,
\end{equation}
where $\vec{v}\in S^{2}$ is a constant unit vector and $\lambda = \tfrac{1}{2}\sqrt{\Lambda}$. In particular, one can perform an $SU(2)_{R}$-rotation to set $\vec{v}=\vec{e}_{3}$. The first equation in \eqref{background} then implies that a $U(1)_{R}$ symmetry is always unbroken in the vacuum, i.e.\ the gauge group in the AdS background is of the form $H\times U(1)_{R}$. Moreover, the second equation shows that this $R$-symmetry is gauged by the graviphoton $\vev{h_{I}}A^{I}_{\mu}$ \cite{Louis:2016qca}.

We are now interested in the moduli space $\cM$ of these vacua. To this end we define the space $\cD$ of scalar deformations $\phi^{i} \rightarrow \vev{\phi^{i}}+\delta \phi^{i}$, $q^{u} \rightarrow \vev{q^{u}}+\delta q^{u}$ that leave \eqref{background} invariant. Moreover we define the space of Goldstone bosons, given by $\cG = \text{span}_{\bbR}\{\vev{k_{I}}\}$.\footnote{This is due to the fact that the kinetic terms for the scalar fields can induce a mass term for the gauge bosons of the form $M_{IJ}\propto \vev{K_{uv}k_{I}^{u}k_{J}^{v}}$ for some invertible matrix $K_{uv}$. See \cite{Louis:2016qca} for an explicit computation.} If the gauge group is spontaneously broken, the corresponding Goldstone bosons are always among the invariant deformations of \eqref{background}, $\cG \subset \cD$. However, they should not be counted as physical moduli. Thus we define the moduli space of the AdS background to be the quotient $\cM = \cD / \cG$. This space is in particular a submanifold of the scalar target space $\cT$. It was shown in \cite{Louis:2016qca} that $\cM \subset \cT_{H}$, i.e.\ all vector multiplet scalars are fixed while scalars from the hypermultiplets can be moduli. Moreover $\cM$ carries a K\"ahler structure that naturally descends from the quaternionic K\"ahler structure of the ambient manifold $\cT_{H}$. Finally, if we denote by $n_{G}= \rk \vev{k_{I}^{u}}$ the number of Goldstone bosons, then the real dimension of the moduli space is
\begin{equation}\label{dimM}
\dim \cM \leq 2n_{H}-2n_{G}\ .
\end{equation}

\section{$\AdS_{5}$ vacua from $T^{1,1}$ compactifications}\label{truncations}

In this section we review consistent truncations of type IIB supergravity compactified on the Sasaki-Einstein manifold $T^{1,1} = \tfrac{SU(2)\times SU(2)}{U(1)}$. Moreover, we study AdS backgrounds which preserve $\cN=2$ supersymmetry in these examples and explicitly compute their moduli spaces. Before we turn to the special case of truncations that preserve $\cN=2$ supersymmetry in five dimensions (8 real supercharges), let us first discuss the general case. Compactifications of type IIB supergravity, their supersymmetry and their consistent truncations have been extensively studied, see for example \cite{Acharya:1998db,Cassani:2010uw,Gauntlett:2010vu,Grimm:2014aha}. Starting with type IIB supergravity, one has 32 real supercharges in ten dimensions. The amount of supersymmetry that is preserved in a consistent truncation to five dimensions depends on the number of linearly independent spinors on the compact manifold. In the case at hand we are interested in compactifications on five-dimensional Sasaki-Einstein manifolds $\SE_{5}$. Such a Sasaki-Einstein manifold admits two linearly independent Killing spinors which are charge conjugate \cite{Sparks:2010sn}, i.e.\ there exists a spinor $\eta$ on $\SE_{5}$ such that
\begin{equation}
 \nabla^{\SE}_{m}\eta = \alpha \gamma_{m}\eta\ , \quad m=1,...,5\ ,
\end{equation}
for some constant $\alpha \in \bbR$. Then the charge conjugate $\eta^{c}$ satisfies the same equation with constant $-\alpha$. This fact was used in \cite{Cassani:2010uw,Gauntlett:2010vu} to prove that a consistent truncation on a Sasaki-Einstein manifold can be described by an $\cN=4$ supergravity in five dimensions, i.e.\ the number of supercharges reduces to 16. One can expand the ten-dimensional gravitino in terms of the spinors on the Sasaki-Einstein manifold as $\Psi^{\cA}= \psi^{\cA}_{1}\otimes \eta \otimes \theta + \psi^{\cA}_{2} \otimes \eta^{c} \otimes \theta^{c}$, giving rise to four symplectic Majorana gravitini. Here $\theta$ is a two-component spinor that appears due to the decomposition of the ten-dimensional Clifford algebra into five-dimensional ones \cite{Gauntlett:2005ww}.

However, we are interested in solutions that contain an $\AdS_{5}$ factor and preserve all supercharges. As discussed in the five-dimensional case above, this imposes $\vev{\delta_{\epsilon} \Psi^{\cA}} = 0$ for $\epsilon$ being the supersymmetry parameter. It was shown in \cite{Gauntlett:2005ww,Liu:2011dw} that for supersymmetric backgrounds of the form $\AdS_{5}\times \SE_{5}$, $\epsilon$ has to be a complex spinor of the form
\begin{equation}
 \epsilon = \psi \otimes \eta \otimes \theta ,
\end{equation}
where $\psi$ is a Killing spinor on $\AdS_{5}$. Since the smallest spinor representations of $Spin(1,4)$ are eight-dimensional, the background only preserves 8 real supercharges, i.e.\ $\cN=2$ supersymmetry in five dimensions. Due to this, one can expect $\AdS_{5}$ solutions of the truncated theory to preserve at most $\cN=2$ supersymmetry. This was indeed shown to be true in \cite{Cassani:2010na}.

The consistent truncations of \cite{Cassani:2010uw,Gauntlett:2010vu} were generalized to include nontrivial second cohomology forms on $T^{1,1}$ in \cite{Cassani:2010na, Bena:2010pr}. There it was shown that the truncated theory can again be interpreted as $\cN=4$, $d=5$ supergravity, coupled to 3 vector multiplets. In particular, these truncations contain an additional $\cN=4$ vector multiplet compared to the truncations discussed in \cite{Cassani:2010uw,Gauntlett:2010vu}, called the $\cN=4$ Betti-vector multiplet because it originates from the fact that $T^{1,1}$ has nontrivial second and third cohomology classes. The reduction then proceeds as follows: type IIB supergravity contains the spacetime metric, the Axion-Dilaton, a doublet of three-forms and the Ramond-Ramond five form. These fields are then expanded in terms of the invariant tensors on $T^{1,1}$ \cite{Cassani:2010na, Bena:2010pr,Liu:2011dw}. Let us describe the five-dimensional field content gained from this expansion: the reduction of the type IIB metric yields the five-dimensional spacetime metric $g_{\mu\nu}$, the graviphoton $A^{0}_{\mu}$, three real scalars $\{u_{1},u_{2},u_{3}\}$ and a complex scalar $v$. Moreover, the reduction of the two three-forms gives two doublets of vectors $\{B_{\mu}^{a}, C_{\mu}^{a}\}$, two doublets of real scalars $\{e^{a},c^{a}\}$ and one doublet of complex scalars $b^{a}$ for $a=1,2$. Finally, the reduction of the RR five-form contributes a complex vector $L_{\mu}$, two real vectors $\{A_{\mu}^{1}, A_{\mu}^{2}\}$ and a real scalar $k$. Combined with the reduction of the Axion-Dilaton $\tau$, the five-dimensional bosonic field content is thus given by the metric $g_{\mu\nu}$, the graviphoton $A_{\mu}^{0}$, eight vectors $\{A^{1}_{\mu}, A^{2}_{\mu}, B_{\mu}^{a}, C_{\mu}^{a},L_{\mu},\bar L_{\mu}\}$, $a=1,2$, and 16 real scalars 
\begin{equation}\label{scalars}
\{u_{1},u_{2},u_{3},c^{a},e^{a},k,\tau,\bar\tau,v,\bar v, b^{a}, \bar b^{a}\}\ .
\end{equation}
The scalar manifold is the coset space
\begin{equation}
\cM_{\text{scalar}} = SO(1,1)\times \frac{SO(5,3)}{SO(5)\times SO(3)}\ .
\end{equation}

Before we proceed, let us discuss the transformation of the five-dimensional fields under the global $SL(2,\bbR)$-symmetry of type IIB supergravity. Since the three-forms $\hat F^{a}_{3}$ in type IIB transform linearly under $SL(2,\bbR)$ \cite{Becker:2007zj}, the same holds for all the five-dimensional fields appearing in their expansion. In particular, for $a,b,c,d\in\bbR$ and $A\in SL(2,\bbR)$ such that
\begin{equation}
A=
\begin{pmatrix}
a & b \\
c & d
\end{pmatrix}\ ,
\quad ad-bc=1\ ,
\end{equation}
an $SL(2,\bbR)$-doublet $V^{a}$ coming from the reduction of the type IIB three-forms transforms
\begin{equation}\label{SL(2,R)}
\begin{pmatrix}
V^{1} \\ V^{2}
\end{pmatrix}
\mapsto 
\begin{pmatrix}
a & b \\
c & d 
\end{pmatrix}
\begin{pmatrix}
V^{1} \\ V^{2}
\end{pmatrix}.
\end{equation}
Moreover, the Axion-Dilaton $\tau$  transforms under the global $SL(2,\bbR)$-symmetry non-linearly via \cite{Becker:2007zj}
\begin{equation}\label{SL(2,R)tau}
\tau \mapsto \frac{a\tau+b}{c\tau+d}\ .
\end{equation}
All remaining fields are invariant under $SL(2,\bbR)$.

Let us now discuss the supersymmetry breaking from $\cN=4$ to $\cN=2$ of this compactification. The $\cN=4$ gravity multiplet decomposes into an $\cN=2$ gravity multiplet, an $\cN=2$ gravitino multiplet and an $\cN=2$ vector multiplet. Moreover, each $\cN=4$ vector multiplet decomposes into an $\cN=2$ vector multiplet and a hypermultiplet, giving in total four vector multiplets and three hypermultiplets. The truncation to $\cN=2$ supergravity is then done by removing the massive $\cN=2$ gravitino multiplet. However, by examining the resulting equations of motion one finds that this alone would not lead to a consistent truncation. In fact, one can show \cite{Cassani:2010na} that for the truncation to be consistent, one additionally has to either truncate the $\cN=2$ Betti-vector multiplet or the Betti-hypermultiplet. These theories were extensively studied in \cite{Halmagyi:2011yd}. There the data we used in section \ref{sugra} to determine the conditions on AdS vacua in a given supergravity were explicitly computed for these $\cN=2$ truncations.

In principle, it would also be consistent to complete truncate the $\cN=4$ Betti multiplet from the spectrum. Since the resulting theories are precisely the ones analyzed in \cite{Cassani:2010uw,Gauntlett:2010vu}, we will not discuss these truncations here. Moreover, reference \cite{Halmagyi:2011yd} also computes moment maps and Killing vectors for the so-called NS truncation, which is done by removing the whole RR sector of type IIB supergravity before reducing to five dimensions. However, this means that the NS truncations does not contain the Axion-Dilaton and thus is not expected to have supersymmetric AdS vacua \cite{Ashmore:2016oug}. We checked this explicitly by inserting the data for the Killing vectors and moment maps from \cite{Halmagyi:2011yd} into equation \eqref{background}. One finds that the resulting equations can not be satisfied and thus no supersymmetric AdS backgrounds exist in this truncation.

In the rest of this section we review the data for the Betti-vector and Betti-hyper truncation \cite{Halmagyi:2011yd} and compute the conditions on AdS backgrounds and their moduli spaces in these five-dimensional theories. Before we proceed, let us collect some differences in notation between this work and \cite{Halmagyi:2011yd}. In general, we replace the indices $i,j=1,2$ with $a,b=1,2$. Moreover, we replaced
\begin{equation}
\begin{aligned}
P_{I}&\rightarrow \mu_{I}\ , \quad b^{i}_{0}\rightarrow b^{a}\ , \quad v_{i} \rightarrow f_{a}\ , \quad A_{1}\rightarrow A_{0}\\
k_{11}&\rightarrow A_{1}\ , \quad k_{12}\rightarrow A_{2}\ , \quad e_{0}^{i}\rightarrow e^{a}\ .
\end{aligned}
\end{equation}

\subsection{The Betti-vector truncation}

We first discuss the Betti-vector truncation, i.e.\ the case where the Betti-hypermultiplet is truncated out of the spectrum. This leads to an $\cN=2$ theory that contains gravity coupled to two vector multiplets and two hypermultiplets \cite{Cassani:2010na}. In this truncation, 10 scalars $\{u_{1},u_{2},u_{3},k,\tau,\bar\tau,b^{a},\bar b^{a}\}$ from \eqref{scalars} are kept. The vector in the gravity multiplet is $A^{0}_{\mu}$ while the other vectors are given by $A^{1,2}_{\mu}$ with associated one-forms $A_{I}$, $I=0,1,2$. The scalars $\{u_{2},u_{3}\}$ of the vector multiplets span the manifold
\begin{equation}
 \cT^{BV}_{V}=SO(1,1)\times SO(1,1)\ .
\end{equation}
The relevant special geometric data is given by \cite{Halmagyi:2011yd}
\begin{equation}\label{BVhup}
 h^{0} = e^{4u_{3}}\, \quad h^{1}=e^{2u_{2}-2u_{3}},\ \quad h^{2}=e^{-2u_{2}-2u_{3}}\ ,
\end{equation}
with $C_{012}=\tfrac{1}{6}$ and all others zero.\footnote{Note that we use conventions for the special real geometry which are different from \cite{Halmagyi:2011yd}, see \cite{Bergshoeff:2004kh} for details.} Lowering the index $I$ according to \eqref{hindex} one finds
\begin{equation}\label{BVhlow}
 h_{0}=\tfrac{1}{3}e^{-4u_{3}}\ , \quad h_{1}=\tfrac{1}{3}e^{-2u_{2}+2u_{3}}\ , \quad h_{2}=\tfrac{1}{3}e^{2u_{2}+2u_{3}}\ .
\end{equation}

The hypermultiplet scalars $\{u_{1},k,\tau,\bar\tau,b^{a}, \bar{b}^{a}\}$ for $a=1,2$ can be shown to span the quaternionic K\"ahler manifold \cite{Cassani:2010na}
\begin{equation}
 \cT_{H}^{BV}=\frac{SO(4,2)}{SO(4)\times SO(2)}\ .
\end{equation}
Note that the scalars $u_{1}$ and $k$ are real while all others are complex. In particular, $\tau = a + ie^{-\phi}$ is the reduction of the Axion-Dilaton of type IIB supergravity. The metric on $\cT^{BV}_{H}$ can be read off from the kinetic terms of the hypermultiplets in the Lagrangian \cite{Halmagyi:2011yd},
\begin{equation}\label{BVL}
\begin{aligned}
 \cL_{Hyper}^{BV} = &-4e^{-4u_{1}+\phi}M_{ab}Db_{0}^{a}\wedge \ast D\bar{b}_{0}^{b}-8du_{1}\wedge\ast du_{1}\\
&-\tfrac{1}{2}e^{-8u_{1}}K\wedge\ast K-\tfrac{1}{2}d\phi \wedge\ast d\phi-\tfrac{1}{2}e^{2\phi}da\wedge\ast da\ ,
\end{aligned}
\end{equation}
where 
\begin{equation}\label{db}
 \begin{aligned}
  Db^{a}&=db^{a}-3ib^{a}A_{0}\ , \\
  K&=Dk+2\epsilon_{ab}[b^{a}D\bar{b}^{b}+\bar{b}^{a}Db^{b}]\ , \\
  Dk&=dk-QA_{0}-2A_{1}-2A_{2}\ ,
  \end{aligned}
\end{equation}
for $Q\in \bbR^{+}$ and
\begin{equation}\label{M}
 M_{ab} = e^{\phi}
 \begin{pmatrix}
 a^{2}+e^{-2\phi} & -a \\
 -a & 1
 \end{pmatrix}\ 
 = \tfrac{1}{\Im \tau}
 \begin{pmatrix}
 |\tau|^{2} & -\Re \tau \\
 -\Re \tau & 1
 \end{pmatrix}\ .
\end{equation}
For $A\in SL(2,\bbR)$, the matrix $M_{ab}$ transforms under the symmetry discussed in \eqref{SL(2,R)} as \cite{Becker:2007zj}
\begin{equation}\label{SL(2,R)M}
M_{ab} \mapsto (A^{-1})_{a}^{c} M_{cd} (A^{-1})_{b}^{d}\ .
\end{equation}

The gauge group is realized on the hypermultiplet scalar manifold via the Killing vectors $k_{I}=k_{I}^{u}\del_{u}$ \cite{Halmagyi:2011yd}
\begin{equation}\label{BVKilling}
\begin{aligned}
 k_{0}&=-3ib^{a}\partial_{b^{a}}+3i\bar{b}^{a}\partial_{\bar{b}^{a}}-Q\partial_{k}\ , \\
 k_{1}&= 2\partial_{k}\ , \\
 k_{2}&= 2\partial_{k}\ . 
\end{aligned}
\end{equation}
The associated moment maps \cite{Halmagyi:2011yd} are\footnote{We converted the moment maps according to $(\mu_{I})_{\cB}^{\cA} = i \vec{\mu}_{I}(\vec{\sigma})^{\cA}_{\cB}$ where $(\vec{\sigma})^{\cA}_{\cB}$ are the Pauli matrices, see \cite{Bergshoeff:2004kh} for more details.}
\begin{equation}\label{BVMoment}
\begin{aligned}
 \vec{\mu}_{0} &= 6e^{-2u_{1}}f_{a}b^{a}\vec{e}_{1}-6e^{-2u_{1}}\bar{f}_{a}\bar{b}^{a}\vec{e}_{2}+(\tfrac{1}{2}e^{-4u_{1}}e^{Z}-3)\vec{e}_{3}\ , \\
 \vec{\mu}_{1} &= -e^{-4u_{1}}\vec{e}_{3}\ , \\
 \vec{\mu}_{2} &= -e^{-4u_{1}}\vec{e}_{3}\ ,
\end{aligned}
\end{equation}
where $e^{Z}=Q-6i\epsilon_{ab}(b^{a}\bar{b}^{b}-\bar{b}^{a}b^{b})$ and $f_{a}$ is defined via $f_{a}b^{a}=\tfrac{1}{\sqrt{\Im \tau}}(b^{2}-\tau b^{1})$. From this we can immediately compute the Lie algebra $\g$ spanned by the $k_{I}$. Since $k_{1}=k_{2}$, we only have to compute the respective Lie brackets with $k_{0}$. We find
\begin{equation}
 [k_{0},k_{1}] =[k_{0},k_{2}] = 0\ ,
\end{equation}
since the vectors $\{ \partial_{a}, \bar{\partial}_{a}, \partial_{k}\}$ are linearly independent and $Q$ is a constant. Thus by observing that the Killing vectors \eqref{BVKilling} have compact and non-compact parts, the gauge group is $G= U(1)\times U(1)\times \bbR$. Note that the scalars are only charged under an $U(1)\times \bbR$ subgroup of $G$ with associated gauge fields $A_{0}$ and $QA_{0}-2A_{1}-2A_{2}$.

To find the AdS vacua of the five-dimensional theory coming from the Betti-vector truncation we have to solve the equations \eqref{background}. For the first equation we use \eqref{BVhlow}, \eqref{BVKilling} and find
\begin{equation}
\vev{h^{I}k_{I}}=-(3i\vev{b^{a}}\del_{b^{a}}-3i\vev{\bar{b}^{a}}\del_{\bar{b}^{a}})e^{4\vev{u_{3}}}+2e^{2\vev{u_{3}}}(e^{2\vev{u_{2}}}+e^{-2\vev{u_{2}}}-\tfrac{Q}{2}e^{6\vev{u_{3}}})\del_{k} = 0\ .
\end{equation}
Due to linear independence of the vectors $\{\del_{b^{a}},\del_{\bar{b}^{a}},\del_{k}\}$, this implies $\vev{b^{a}}=\vev{\bar{b}^{a}}=0$ and
\begin{equation}\label{BVu2}
e^{2\vev{u_{2}}}+e^{-2\vev{u_{2}}}=\tfrac{Q}{2}e^{6\vev{u_{3}}}\ .
\end{equation}
In particular, the background values of the vector multiplet scalars are not independent of each other. Inserting these results into the moment maps \eqref{BVMoment}, we find that only the third component $\vev{\mu_{I}}:=\vev{\mu_{I}^{3}}$ is nonzero in the background. These components are given by
\begin{equation}
\vev{\mu_{0}}=\tfrac{Q}{2}e^{-4\vev{u_{1}}}-3\ , \quad \vev{\mu_{1}}=\vev{\mu_{2}}=-e^{-4\vev{u_{1}}}\ ,
\end{equation}
where we used the fact that $\vev{e^{Z}}=Q$ in the AdS background.

We are thus left with solving the second equation in \eqref{background}. Since $\vev{\mu_{1}}=\vev{\mu_{2}}$, \eqref{background} implies that $\vev{h_{1}}=\vev{h_{2}}$ and thus using \eqref{BHhlow} we find
\begin{equation}
e^{-2\vev{u_{2}}+2\vev{u_{3}}}=e^{2\vev{u_{2}}+2\vev{u_{3}}}\ .
\end{equation}
This fixes $\vev{u_{2}}=0$. Also, by using \eqref{BVu2} we find $\vev{u_{3}}=\tfrac{1}{6}\log \tfrac{4}{Q}$. Note that this fixes $\vev{u_{3}}$, since $Q$ is a constant. Now consider the zero-component $\vev{\mu_{0}}$. Using again \eqref{BVhlow} and \eqref{BVMoment} and inserting this into the second equation of \eqref{background}, we find
\begin{equation}
\tfrac{Q}{2}e^{-\vev{u_{1}}}-3=\tfrac{\lambda}{3}e^{-4\vev{u_{3}}}=\tfrac{Q^{2/3}\lambda}{6\sqrt[3]{2}}\ ,
\end{equation}
fixing the background value of the scalar $\vev{u_{1}}=-\log(\tfrac{\lambda}{3\sqrt[3]{2}Q^{1/3}}+ \tfrac{6}{Q})$. To summarize, the conditions for $\AdS_{5}$ vacua from the Betti-vector truncation fix all the scalars $\{u_{2},u_{3}\}$ from the vector multiplets and moreover the scalars $\{u_{1},b^{a},\bar{b}^{a}\}$ from the hypermultiplets.

Let us now turn to the moduli space $\cM^{BV}$ described in section \ref{sugra}. From the above analysis we know that the hypermultiplet scalars $\{k,\tau,\bar\tau\}$ are not constrained by the conditions \eqref{background} for the AdS background. However, the moduli space was proven to be a K\"ahler manifold and thus has to be even-dimensional. This can be understood by considering the background values of the Killing vectors \eqref{BVKilling},
\begin{equation}
\vev{k_{0}}=-Q\del_{k}\ , \quad \vev{k_{1}}=\vev{k_{2}}=2\del_{k}\ .
\end{equation}
From this we see that the space of Goldstone bosons is one-dimensional and spanned by $\del_{k}$. The respective scalar $k$ then gets eaten by the vector field $QA_{0}-2A_{1}-2A_{2}$, which becomes massive as a result of the symmetry breaking. In particular, we explained in section \ref{sugra} that the $U(1)_{R}$ symmetry always remains unbroken in the vacuum and is gauged by the graviphoton 
\begin{equation}
\vev{h_{I}}A^{I}_{\mu}= \tfrac{Q^{2/3}}{6\sqrt[3]{2}}A_{\mu}^{0}+\tfrac{2^{2/3}}{3\sqrt[3]{Q}}(A^{1}_{\mu}+A^{2}_{\mu})\ . 
\end{equation}
Thus we find that the gauge group in the supersymmetric AdS background is broken according to $U(1)\times U(1)\times \bbR\longrightarrow U(1)\times U(1)_{R}$. 

We have shown that the moduli space of the AdS vacuum is two-dimensional and spanned by the Axion-Dilaton $\{\tau, \bar\tau\}$. This agrees with the bound \eqref{dimM} on the dimension of the moduli space for $n_{H}=2$, $n_{G}=1$,
\begin{equation}
\dim \cM^{BV}\leq 2\cdot 2-2\cdot 1= 2\ ,
\end{equation}
i.e.\ the moduli space is of maximal dimension. To compute the metric $g^{BV}$ on the moduli space, note that the coordinate one-forms of the fixed scalars vanish on $\cM^{BV}$, i.e.\
\begin{equation}
db^{a}|_{\cM^{BV}}=d\bar{b}^{a}|_{\cM^{BV}}=du_{1}|_{\cM^{BV}}=dk|_{\cM^{BV}}=0\ .
\end{equation}
Using this fact we can read off the metric from the Lagrangian \eqref{BVL},
\begin{equation}\label{upperhalfplane}
g^{BV}= d\phi^{2}+e^{2\phi}da^{2} = \tfrac{1}{\Im \tau^{2}}d\tau d\bar\tau\ .
\end{equation}
This is precisely the metric on the upper half plane $\cH$ which is well-known to be a K\"ahler manifold.  Thus the moduli space of the $\AdS_{5}$ vacua in the Betti-vector truncation is
\begin{equation}\label{BVmetric}
\cM^{BV}= \cH \ .
\end{equation}

\subsection{The Betti-hyper truncation}

Let us now turn to the Betti-hyper truncation from which one obtains a five-dimensional $\cN=2$ theory that contains gravity coupled to one vector multiplet and three hypermultiplets \cite{Cassani:2010na}. Here, 13 scalars $\{u_{1},u_{3},k,e^{a},\tau,\bar\tau,v,\bar v, b^{a}, \bar b^{a}\}$ are kept from \eqref{scalars}. While the graviphoton is still $A^{0}_{\mu}$, the vector in the vector multiplet is given as the combination $\tfrac{1}{2}(A^{1}_{\mu}+A^{2}_{\mu})$. The single scalar $u_{3}$ in the vector multiplet parametrizes the manifold
\begin{equation}
\cT_{V}^{BH}=SO(1,1)\ .
\end{equation}
Its special real geometry is given by \cite{Halmagyi:2011yd}
\begin{equation}\label{BHhup}
h^{0}=e^{4u_{3}}\ , \quad h^{1}=e^{-2u_{3}}\ ,
\end{equation}
with $C_{011}=\tfrac{1}{3}$ and all others zero. Moreover we obtain
\begin{equation}\label{BHhlow}
h_{0}=\tfrac{1}{3}e^{-4u_{3}}\ , \quad h_{1}=\tfrac{2}{3}e^{2u_{3}}\ ,
\end{equation}
by lowering the index according to \eqref{hindex}.

In the hypermultiplets, the scalars $\{ u_{1}, k, e^{a}, b^{a}, \bar{b}^{a},\tau,\bar\tau, v, \bar v\}$ span the quaternionic K\"ahler manifold \cite{Cassani:2010na}
\begin{equation}
\cT^{BH}_{H}=\frac{SO(4,3)}{SO(4)\times SO(3)}\ .
\end{equation}
Here $b^{a}$ and $v$ are complex while all others are real.\footnote{We replace $(e_{1},e_{2})\mapsto (-e_{2},e_{1})$ compared to \cite{Halmagyi:2011yd}.} The hypermultiplet Lagrangian is given by \cite{Halmagyi:2011yd}
\begin{equation}\label{BHL}
\begin{aligned}
\cL_{Hyper}^{BH}=&-e^{-4u_{1}}M_{ab}[\tfrac{1}{2}De^{a}\wedge\ast De^{b}+\tfrac{1}{2}G^{a}\wedge\ast G^{b}+2(F^{a}\wedge\ast \bar{F}^{b}+\bar{F}^{a}\wedge\ast F^{b})]\\
&-8du_{1}\wedge\ast du_{1}-d|v|\wedge\ast d|v|+|v|^{2}D\theta \wedge\ast D\theta\\
&-\tfrac{1}{2}e^{-8u_{1}}K\wedge\ast K-\tfrac{1}{2}d\phi\wedge\ast d\phi-\tfrac{1}{2}e^{2\phi}da\wedge\ast da\ ,
\end{aligned}
\end{equation}
where
\begin{equation}
\begin{aligned}
G^{a}&=(1+|v|^{2})De^{a}-4\Im(vDb^{a})\ , \\
F^{a}&=Db^{a}-\tfrac{i}{2}\bar{v}De^{a}\ , \\
De^{a}&=de^{a}-j^{a}A_{0}\ , \\
D\theta &= d\theta+3A_{0}\ ,
\end{aligned}
\end{equation}
$j^{a}$ are constant charges and $\theta=\theta(v,\bar v)$ is a function of $v$ and $\bar v$ whose explicit form is not important for our purposes. Moreover $Db^{a}$ and $M_{ab}$ are defined in \eqref{db} and \eqref{M}, respectively. From this Lagrangian one can read off the metric for the hypermultiplet scalars.

The gauge group acts on $\cT_{H}^{BH}$ via the Killing vectors \cite{Halmagyi:2011yd}
\begin{equation}\label{BHKilling}
\begin{aligned}
k_{0}&=-(Q+\epsilon_{ab}j^{a}e^{b})\del_{k}-3ib^{a}\del_{b^{a}}+3i\bar{b}^{a}\del_{\bar{b}^{a}}+\tfrac{3}{2}(1+\rho^{2})\del_{\rho}+\tfrac{3}{2}(1+\bar{\rho}^{2})\del_{\bar\rho}-j^{a}\del_{e^{a}}\ , \\
k_{1}&=4\del_{k}\ ,
\end{aligned}
\end{equation}
where we introduced a complex scalar $\rho\in \cH$ in the upper half plane that is related to $v$ via
\begin{equation}
v=-\frac{(i-\rho)(i-\bar\rho)}{1+|\rho|^{2}}\ .
\end{equation}
The associated moment maps are then given by \cite{Halmagyi:2011yd}
\begin{equation}\label{BHMoment}
 \begin{aligned}
  \vec{\mu}_{0} & =  (\tfrac{1}{2}e^{\tilde Z}e^{-4u_{1}}-\tfrac{3}{2\Im \rho}(1+\lvert \rho \rvert ^{2}))\vec{e}_{3} \\
  &-\tfrac{1}{2\rho}e^{-2u_{1}}f_{a}(-3i(\bar{\rho}-i)^{2}\bar{b}^{a}-3i(\bar{\rho}+i)^{2}b^{a}+i(1-i\bar{\rho})(1+i\bar{\rho})j^{a})\vec{e_{1}} \\
  &+\tfrac{1}{2\rho}e^{-2u_{1}}\bar{f}_{a}(3i(\rho-i)^{2}b^{a}+3i(\rho+i)^{2}\bar{b}^{a}+i(1-i\rho)(1+i\rho)j^{a})\vec{e_{2}}\ , \\
  \vec{\mu}_{1}&=-2e^{-4u_{1}}\vec{e}_{3}\ ,
 \end{aligned}
\end{equation}
where $e^{\tilde Z}=e^{Z}+\epsilon_{ab}(j^{a}e^{b}-e^{a}j^{b})= Q-6i\epsilon_{ab}(b^{a}\bar{b}^{b}-\bar{b}^{a}b^{b})+\epsilon_{ab}(j^{a}e^{b}-j^{b}e^{a})$. Since $[k_{0},k_{1}]=0$, the gauge group in this case is $G=U(1)\times \bbR$.

Let us now solve \eqref{background} in this compactification. To do this, we insert \eqref{BHKilling} and \eqref{BHhup} into the first equation of \eqref{background},
\begin{equation}
\begin{aligned}
 0=\vev{h^{I}k_{I}}=(&4e^{-6\vev{u_{3}}}-Q-\epsilon_{ab}j^{a}\vev{e^{b}})e^{4\vev{u_{3}}}\del_{k} -(3i\vev{b^{a}}\del_{b^{a}}-3i\vev{\bar b^{a}}\del_{\bar b^{a}})e^{4\vev{u_{3}}}\\
 &+\tfrac{3}{2}(1+\vev{\rho^{2}})e^{4\vev{u_{3}}}\del_{\rho}+\tfrac{3}{2}(1+\vev{\bar\rho^{2}})e^{4\vev{u_{3}}}\del_{\bar\rho}\ .
\end{aligned}
 \end{equation}
Using the linear independence of the basis vectors $\{\del_{b^{a}}, \del_{\bar{b}^{a}}, \del_{k}, \del_{\rho}, \del_{\bar\rho}\}$ we immediately find $\vev{b^{a}}=\vev{\bar{b}^{a}}=j^{a}=0$.\footnote{The vanishing of the topological charges $j^{a}$ shows that the backgrounds we are discussing are indeed related to the Klebanov-Witten theory \cite{Klebanov:1998hh} and not the Klebanov-Strassler solutions \cite{Klebanov:2000hb}.} Moreover, we learn that
\begin{equation}
\vev{\rho^{2}}=\vev{\bar{\rho}^{2}}=-1\ , \quad Q=4e^{-6\vev{u_{3}}}\ ,
\end{equation}
where we used $j^{a}=0$. In particular, we find $\vev{e^{\tilde{Z}}}=Q$ and $\vev{u_{3}}=-\tfrac{1}{6}\log\tfrac{Q}{4}$. The first equation is solved by $\vev{\rho}=i$ which implies $\vev{v}=\vev{\bar v}=0$ and 
\begin{equation}
\vev{|\rho|^{2}}=1\ .
\end{equation}
Thus the vector multiplet scalar $u_{3}$ and the hypermultiplet scalars $\{b^{a},\bar{b}^{a},v,\bar v\}$ are fixed by the first equation in \eqref{background}. Using the above results, we find that the moment maps \eqref{BHMoment} are only nontrivial in the $\vec{e}_{3}$-direction and read
\begin{equation}
\vev{\mu_{0}}=\tfrac{Q}{2}e^{-4\vev{u_{1}}}-3\ , \quad \vev{\mu_{1}}=-2e^{-4\vev{u_{1}}}\ .
\end{equation} 
Inserting these expressions into the second equation of \eqref{background} and using \eqref{BHhlow}, we find
\begin{equation}
(\tfrac{Q}{2}e^{-4\vev{u_{1}}}-3)=3\lambda e^{4\vev{u_{3}}}\ , \quad e^{-4\vev{u_{1}}}=\tfrac{2}{3}\lambda e^{-2\vev{u_{3}}}\ .
\end{equation}
These equations fix the scalar $\vev{u_{1}}=-\tfrac{1}{4}\log(\tfrac{2^{5/3}\lambda}{3\sqrt[3]{Q}})$. In conclusion, we have shown that the $\AdS_{5}$ conditions for the Betti-hyper truncation fix the vector multiplet scalar $u_{3}$ and the hypermultiplet scalars $\{b^{a},\bar{b}^{a},v,\bar v, u_{1}\}$.

To compute the moduli space, we observe that the scalars $\{k, e^{a},\tau, \bar\tau\}$ are not restricted by the AdS conditions \eqref{background} and thus their associated deformations leave the vacuum invariant. However, as before we find that the Killing vectors have nontrivial background values,
\begin{equation}
\vev{k_{0}}=-Q\del_{k}\ , \quad \vev{k_{1}}=4\del_{k}\ ,
\end{equation}
and thus the space of Goldstone bosons is again one-dimensional. The gauge group is broken, $U(1)\times \bbR\longrightarrow U(1)_{R}$, and the vector field $QA_{0}-2(A_{1}+A_{2})$ becomes massive by ``eating" the scalar $k$. Note that, as discussed in section \ref{sugra}, the $U(1)_{R}$ symmetry of the background is still present after the spontaneous symmetry breaking. It is gauged by the graviphoton
\begin{equation}
\vev{h_{0}}A^{0}_{\mu}+\tfrac{1}{2}\vev{h_{1}}(A_{\mu}^{1}+A_{\mu}^{2})= \tfrac{Q^{2/3}}{6\sqrt[3]{2}}A_{\mu}^{0}+\tfrac{2^{2/3}}{3\sqrt[3]{Q}}(A^{1}_{\mu}+A^{2}_{\mu})\ . 
\end{equation}
The moduli space is thus four-dimensional and spanned by the hypermultiplet scalars $\{e^{a}, \tau, \bar\tau\}$. This is in agreement with the bound \eqref{dimM} for $n_H=3$, $n_{G}=1$,
\begin{equation}
\dim \cM^{BH}\leq 2\cdot 3 - 2\cdot 1 =4\ . 
\end{equation}
Note that $\dim \cM^{BH}=4$ is again the maximal dimension possible for the given number of hypermultiplets and Goldstone bosons.

The moduli space is thus spanned by the reduction of the Axion-Dilaton $\tau$ and a doublet of real scalars $e^{a}$ coming from the Betti-hypermultiplet. To compute the metric, we use the fact that the coordinate one-forms of the fixed scalars vanish on $\cM^{BH}$. Inserting this into the Lagrangian  \eqref{BHL} for the hypermultiplets, we find
\begin{equation}\label{metric}
g^{BH}= \gamma M_{ab}de^{a}de^{b}+\tfrac{1}{\Im\tau^{2}}d\tau d\bar\tau\ ,
\end{equation}
where $\gamma = 2e^{-4\vev{u_{1}}}=\tfrac{2}{3}(2Q)^{1/3}\lambda$ and $M_{ab}$ is defined in \eqref{M}. We immediately recognize the second term in \eqref{metric} as the metric of the Axion-Dilaton on the upper half plane \eqref{upperhalfplane}. Let us first discuss the isometries of the metric \eqref{metric}. Clearly, \eqref{metric} is invariant under shifts in the scalars $e^{a}$, i.e.\ $e^{a}\mapsto e^{a}+w^{a}$ for some constants $w^{a}$. Moreover, the metric has an $SL(2,\bbR)$-isometry induced by the global $SL(2,\bbR)$-symmetry of type IIB supergravity \cite{Becker:2007zj}. For the term $M_{ab}de^{a}de^{b}$, this follows from the fact that the transformations \eqref{SL(2,R)} and \eqref{SL(2,R)M} exactly cancel each other. Moreover, the metric $\tfrac{1}{\Im\tau^{2}}d\tau d\bar\tau$ on the upper half plane is known to have an $SL(2,\bbR)$-isometry given by \eqref{SL(2,R)tau}.  Thus the metric \eqref{metric} has an $\bbR^{2}\times SL(2,\bbR)$ isometry group. 

We already discussed that the moduli space of $\AdS_{5}$ vacua should be K\"ahler and in particular complex. To this end, let us define a complex structure for the scalars $\{ e^{a}, a, \phi\}$ and construct the K\"ahler potential associated to the metric $g^{BH}$, i.e.\ a real function $K$ such that $(g^{BH})_{i\bar j} = \del_{i}\del_{\bar j} K$ for $i,j =1,2$ complex indices on $\cM^{BH}$. For the scalars $\{a,\phi\}$, the complex structure is naturally given by the Axion-Dilaton $\tau = a +ie^{-\phi}$. To define a complex structure on the scalars $e^{a}$, recall that the three-forms $\hat F^{a}_{3}$ in type IIB supergravity can be combined into a complex three-form \cite{Becker:2007zj},
\begin{equation}
\hat G_{3}:= \hat F_{3}^{2} - \tau \hat F_{3}^{1}\ .
\end{equation}
Translating this to the scalars $e^{a}$, we may define a complex scalar $z$ by
\begin{equation}\label{z}
z:=e^{2}-\tau e^{1}\ , \quad \bar z := e^{2}-\bar\tau e^{1}\ .
\end{equation}
In particular, this implies $e^{1}=-\tfrac{\Im z}{\Im \tau}$. The associated coordinate one-forms then introduce a twist between $z$ and $\tau$,
\begin{equation}
dz=de^{2}-\tau de^{1}-e^{1}d\tau\ , \quad d\bar z = de^{2}-\bar\tau de^{1}-e^{1}d\bar\tau\ .
\end{equation}
Using these we can rewrite the metric of the $e^{a}$ in terms of $z$ and $\bar z$,
\begin{equation}
M_{ab}de^{a}de^{b}=\frac{\Im z^{2}}{\Im \tau ^{3}}d\tau d\bar\tau + \frac{1}{\Im \tau}dzd\bar z -\frac{\Im z}{\Im \tau^{2}}(d\tau d\bar z + d\bar\tau dz)\ .
\end{equation}
Thus the full complex metric reads
\begin{equation}\label{BHmetric}
g^{BH}=\left(\frac{1}{\Im\tau^{2}}+\gamma \frac{\Im z^{2}}{\Im\tau^{3}}\right)d\tau d\bar\tau - \gamma \frac{\Im z}{\Im\tau^{2}}(d\tau d\bar z +d\bar\tau dz)+\frac{\gamma}{\Im\tau}dzd\bar z\ .
\end{equation}
This metric is derived from the K\"ahler potential
\begin{equation}\label{Kahlerpotential}
K=-4\log(\tau-\bar\tau)-i\gamma\tfrac{(z-\bar z)^{2}}{\tau-\bar\tau}\ ,
\end{equation}
and the associated K\"ahler form $\omega = \tfrac{i}{2}\del \bar\del K$ is closed. Thus the moduli space is a K\"ahler manifold with K\"ahler structure defined by $K$.

Let us now show that \eqref{BHmetric} extends to a globally well-defined metric and identify the manifold $\cM^{BH}$. Examining the metric \eqref{metric}, we already observed that the second part is the metric $\tfrac{1}{\Im \tau^{2}}d\tau d\bar\tau$ on the upper half plane $\cH$. The first term in \eqref{metric} is the metric of a torus $\bbC/\Lambda_{\tau}$ with complex structure parameter $\tau$. Here $\Lambda_{\tau} = \bbZ\oplus\tau\bbZ$ is a lattice spanned by $(1,\tau)$. However, this description only holds locally. Globally, the moduli space is not a direct product of a complex torus with the upper half plane, since the complex structure \eqref{z} on the torus varies with $\tau$. Thus the global metric is the metric on the total space of a complex torus bundle over the upper half plane, i.e.
\begin{equation}\label{fibration}
\bbC/\Lambda_{\tau}\hookrightarrow \cM^{BH} \longrightarrow \cH\ ,
\end{equation}
Note that this agrees with the results stated in \cite{Klebanov:1998hh}; the moduli are the Axion-Dilaton and a complex scalar parametrizing a torus. However, it turns out that the metric is in general not a product metric but the metric of a non-trivial fibration. To identify the total space of the fibration \eqref{fibration}, consider the universal elliptic curve $\cE$ over the upper half plane.\footnote{For an introduction to elliptic curves, their moduli spaces and the universal elliptic curve, see \cite{RHain}.} This is defined as the quotient
\begin{equation}
\cE = \left(\bbC \times \cH\right)/\bbZ^{2}\ ,
\end{equation}
where $(m,n)\in\bbZ^{2}$ acts as
\begin{equation}\label{Z2action}
(z,\tau) \mapsto (z+m+n\tau,\tau)\ .
\end{equation}
Since this action is free and proper, the quotient $\cE$ is a two-dimensional complex manifold \cite{RHain}. In particular, the fibers of the projection $\cE \rightarrow \cH$ are precisely the complex tori $\bbC/\Lambda_{\tau}$. To see that $g^{BH}$ gives a well-defined metric on $\cE$, we have to show that it is compatible with the quotient by the action \eqref{Z2action}. Since $\tau$ is fixed by \eqref{Z2action}, only the second term in the K\"ahler potential \eqref{Kahlerpotential} transforms non-trivially. In particular, we find for the transformation of the K\"ahler potential,
\begin{equation}
K \mapsto K'=K + 2i\gamma n(z-\bar z) + i\gamma n^{2}(\tau-\bar\tau)\ ,
\end{equation}
which is just a K\"ahler transformation $K \mapsto K' = K + f(\tau,z) + \bar f(\bar \tau, \bar z)$ for a holomorphic function $f(\tau,z) = 2i\gamma nz +i\gamma n^{2}\tau$. Thus both potentials give rise to the same K\"ahler metric and $g^{BH}$ is a well-defined global metric on $\cE$.\footnote{A different way to see the invariance of the metric under the $\bbZ^{2}$-action is to realize \eqref{Z2action} via the $\bbR^{2}$-isometry of the metric \eqref{metric} by $e^{1}\mapsto e^{1}-n$, $e^{2}\mapsto e^{2}+m$.} In conclusion, the moduli space of AdS vacua in the Betti-hyper truncation is given by the total space of the universal elliptic curve,
\begin{equation}
\cM^{BH}=\cE = (\bbC\times \cH)/\bbZ^{2}\ .
\end{equation}
This manifold is in particular a homogeneous space, since it has a transitive group action given by the isometries of the metric \eqref{metric}, $\bbR^{2}\times SL(2,\bbR)\cong \bbC \times SL(2,\bbR)$. Because the upper half plane $\cH$ can be written as the quotient $SL(2,\bbR)/SO(2)$, we find
\begin{equation}
\cM^{BH} = [\bbC \times SL(2,\bbR)/SO(2)]/\bbZ^{2}\ .
\end{equation}
Thus for both truncations with homogeneous scalar target spaces, also the AdS moduli space is a homogeneous space.

Before we conclude this section, let us briefly note the following: consider the K\"ahler potential $L$ of $SU(2,1)/U(2)$, given by
\begin{equation}
L = -\log(\tau-\bar\tau + i\epsilon(z-\bar z)^{2})\ ,
\end{equation}
where $\epsilon \in \bbR$ is a constant. To make contact with the potential \eqref{Kahlerpotential}, we want to split off a term of the form $\log(\tau-\bar\tau)$ from $L$. To this end, we separate a factor $\tau-\bar\tau$ inside the logarithm,
\begin{equation}
L=-\log(\tau-\bar\tau + i\epsilon(z-\bar z)^{2})=-\log(\tau-\bar\tau)-\log(1+i\epsilon \tfrac{(z-\bar z)^{2}}{\tau-\bar\tau})\ .
\end{equation}
For small $\epsilon$ the second term can be expanded,
\begin{equation}
-\log(1+i\epsilon\tfrac{(z-\bar z)^{2}}{\tau-\bar\tau})\simeq -i\epsilon\tfrac{(z-\bar z)^{2}}{\tau-\bar\tau} + \cO(\epsilon^{2})\ .
\end{equation}
Thus we can write
\begin{equation}
L=-\log(\tau-\bar\tau +i\epsilon (z-\bar z)^{2})\simeq \tfrac{1}{4}K + \cO(\epsilon^{2})\ ,
\end{equation}
for suitable $\epsilon$ and find that the K\"ahler potential \eqref{Kahlerpotential} of the AdS moduli space appears as a first order term in the $\epsilon$-expansion. To interpret this result we first note the following: we can reinstall the five-dimensional gravitational constant $\kappa$ into the metric \eqref{BHmetric} by $\gamma \mapsto \kappa^{2} \gamma$. Thus the $\epsilon$-expansion performed above actually corresponds to an expansion in the gravitational constant $\kappa$ for $\epsilon=4\kappa^{2}\gamma$ and fixed $\gamma$. Since the limit $\kappa\rightarrow 0$ corresponds to the large $N$ limit of the dual field theory\footnote{The AdS/CFT correspondence relates $\kappa \propto 1/N$, see \cite{Polchinski:2010hw} for a review.}, we can interpret the metric $g_{\cM}^{BH}$ on the moduli space as the first order contribution in a large $N$ expansion of the metric on $SU(2,1)/U(2)$.

\section{Conclusions}\label{conclusion}

Let us summarize the results obtained. We computed the conditions on AdS vacua in five-dimensional $\cN=2$ supergravity for both the $\cN=2$ Betti-vector truncation and the Betti-hyper truncation coming from ten-dimensional type IIB supergravity compactified on $T^{1,1}$. Since these truncations are consistent, our results lift to solutions of the full ten-dimensional supergravity. We find that both truncations admit $\AdS_{5}$ backgrounds which retain an $U(1)_{R}$ symmetry in the vacuum. This is in agreement with the $U(1)_{R}$ symmetry coming from the dual field theories predicted by the AdS/CFT correspondence \cite{Maldacena:1997re,Morrison:1998cs,Klebanov:1998hh}.

The moduli space of the full ten-dimensional solution $\AdS_{5}\times T^{1,1}$ is known to be complex five-dimensional \cite{Klebanov:1998hh,Benvenuti:2005wi,Ardehali:2014zfa,Ashmore:2016qvs,Ashmore:2016oug}. However, only two of those moduli transform as singlets under the $SU(2)\times SU(2)$ factor in the isometry group and are thus accessible via consistent truncations \cite{Hoxha:2000jf}; those are the Axion-Dilaton $\tau$ and the complex modulus $z$ related to the topology of $T^{1,1}$, i.e.\ coming from the fact that $b_{2}(T^{1,1})=1$. In \cite{Ashmore:2016qvs,Ashmore:2016oug} the moduli spaces for type IIB solutions of the form $\AdS_{5}\times \SE_{5}$ were computed from generalized geometry. In particular, it was shown that the Axion-Dilaton is always a modulus, independent from the topology of the Sasaki-Einstein manifold used for the compactification. Moreover, it was shown in \cite{Louis:2016qca} that the moduli of five-dimensional AdS backgrounds must always be recruited out of the hypermultiplets. Our present results agree with these predictions; the moduli space of the Betti-vector truncation is spanned only by the Axion-Dilaton residing in one of the hypermultiplets and the metric is the expected one on the upper half-plane. This can be understood as follows: the fact that $T^{1,1}$ has nontrivial second cohomology leads to the presence of an additional two-form in the reduction ansatz \cite{Cassani:2010na, Bena:2010pr}. This additional two-form gives rise to an additional $\cN=4$ vector multiplet which splits into an $\cN=2$ vector multiplet and an $\cN=2$ hypermultiplet. However, only one of the two can be retained in a consistent $\cN=2$ truncation \cite{Cassani:2010na}. Thus in the Betti-vector truncation the only hypermultiplet related to the topology of $T^{1,1}\cong S^{2}\times S^{3}$ was removed and one would not expect to find the modulus $z$ in this truncation in the first place.

In the case of the Betti-hyper truncation the situation is different. Here the topology of $T^{1,1}$ contributes to the five-dimensional hypermultiplets of the truncation and gives rise to an additional complex modulus $z$. This modulus parametrizes a complex torus with complex structure parameter given by the Axion-Dilaton $\tau$. Thus the complex two-dimensional moduli space $\cM^{BH}$ of the Betti-hyper truncation contains all the moduli of $\AdS_{5}\times T^{1,1}$ that are expected to arise in a consistent truncation. However, even though the Axion-Dilaton is completely unrelated to the geometry of the compact Sasaki-Einstein manifold, it turns out that the metric on the moduli space $\cM^{BH}$ is not a product metric with the Axion-Dilaton split off from the moduli corresponding to the geometry. Indeed, we find that $\cM^{BH}$ is a torus bundle with base space given by the upper half-plane $\cH$. We showed that this bundle is the universal elliptic curve $\cE\rightarrow \cH$ with total space given by $\cE=(\bbC\times \cH)/\bbZ^{2}$. Thus since the Betti-hyper truncation is consistent and lifts to the full ten-dimensional supergravity, the AdS/CFT correspondence relates this result to the metric on a submanifold of the conformal manifold of the Klebanov-Witten theory in four dimensions.\footnote{Consistency of the truncation is related to closure of the associated operators under the operator product expansion in the dual CFT. This then defines the submanifold of the conformal manifold dual to the moduli spaces we discuss here.} It would be interesting to check this result on the field theory side.

Since $T^{1,1}$ is a member of the infinite family of Sasaki-Einstein manifolds called $Y^{p,q}$ \cite{Gauntlett:2004yd}, a natural question would be to extend our computations to these manifolds. On first glance, this might seem easily possible; $Y^{p,q}\cong S^{2}\times S^{3}$ have the same topology as $T^{1,1}$ and a reduced isometry algebra $\mathfrak{su}(2)\oplus\mathfrak{u}(1)\oplus\mathfrak{u}(1)$. Moreover the moduli of the ten-dimensional solutions $\AdS_{5}\times Y^{p,q}$ also contain the Axion-Dilaton and the modulus $z$ from the VEV of the B-field integrated over the nontrivial two-cycle. Additionally, there only exists one other complex modulus transforming in a triplet under the $SU(2)$ in the isometry group, making the full moduli space complex three-dimensional. However, the $Y^{p,q}$ manifolds are in general not coset spaces but only admit a cohomgeneity-one action of the isometry group. Since the techniques used in \cite{Cassani:2010na,Bena:2010pr} rely heavily on the transitivity of the $SU(2)\times SU(2)$ action on the $T^{1,1}$ coset, a similar reduction for the $Y^{p,q}$ manifolds might not be consistent.

\section*{Acknowledgments}

This work was supported by the German Science Foundation (DFG) under
the Collaborative Research Center (SFB) 676 ``Particles, Strings and the Early
Universe'', the Research Training Group (RTG) 1670 ``Mathematics
inspired by String Theory and Quantum Field Theory'' and the Joachim-Herz Stiftung (JHS).

We have benefited from conversations and correspondence with Peter-Simon Dieterich, Jonathan Fisher, Alexander Haupt, Carsten Liese, Severin L\"ust and Daniel Waldram.





\newpage 


\begin{thebibliography}{10}

\bibitem{Gauntlett:2005ww}
  J.~P.~Gauntlett, D.~Martelli, J.~Sparks and D.~Waldram,
  ``Supersymmetric AdS(5) solutions of type IIB supergravity,''
  Class.\ Quant.\ Grav.\  {\bf 23} (2006) 4693
  [hep-th/0510125].
  
\bibitem{Gabella:2009wu}
  M.~Gabella, J.~P.~Gauntlett, E.~Palti, J.~Sparks and D.~Waldram,
  ``AdS(5) Solutions of Type IIB Supergravity and Generalized Complex Geometry,''
  Commun.\ Math.\ Phys.\  {\bf 299} (2010) 365
  [arXiv:0906.4109 [hep-th]].

\bibitem{Coimbra:2015nha}
  A.~Coimbra and C.~Strickland-Constable,
  ``Generalised Structures for $\mathcal{N}=1$ AdS Backgrounds,''
  arXiv:1504.02465 [hep-th].
  
\bibitem{Ashmore:2016qvs}
  A.~Ashmore, M.~Petrini and D.~Waldram,
  ``The exceptional generalised geometry of supersymmetric AdS flux backgrounds,''
  arXiv:1602.02158 [hep-th].

\bibitem{Ashmore:2016oug}
  A.~Ashmore, M.~Gabella, M.~Gra\~na, M.~Petrini and D.~Waldram,
  ``Exactly marginal deformations from exceptional generalised geometry,''
  arXiv:1605.05730 [hep-th].

  
\bibitem{Grana:2016dyl}
  M.~Gra\~na and P.~Ntokos,
  ``Generalized geometric vacua with eight supercharges,''
  JHEP {\bf 1608} (2016) 107
  [arXiv:1605.06383 [hep-th]].


\bibitem{Maldacena:1997re}
  J.~M.~Maldacena,
  ``The Large N limit of superconformal field theories and supergravity,''
  Int.\ J.\ Theor.\ Phys.\  {\bf 38} (1999) 1113
   [Adv.\ Theor.\ Math.\ Phys.\  {\bf 2} (1998) 231]
   [hep-th/9711200].

\bibitem{Morrison:1998cs}
  D.~R.~Morrison and M.~R.~Plesser,
  ``Nonspherical horizons. 1.,''
  Adv.\ Theor.\ Math.\ Phys.\  {\bf 3} (1999) 1
  [hep-th/9810201].

\bibitem{Polchinski:2010hw}
  J.~Polchinski,
  ``Introduction to Gauge/Gravity Duality,''
  arXiv:1010.6134 [hep-th].


\bibitem{Tachikawa:2005tq}
  Y.~Tachikawa,
  ``Five-dimensional supergravity dual of a-maximization,''
  Nucl.\ Phys.\ B {\bf 733} (2006) 188
  [hep-th/0507057].

  
\bibitem{deAlwis:2013jaa}
  S.~de Alwis, J.~Louis, L.~McAllister, H.~Triendl and A.~Westphal,
  ``Moduli spaces in $AdS_4$ supergravity,''
  JHEP {\bf 1405} (2014) 102
  [arXiv:1312.5659 [hep-th]].

\bibitem{Louis:2014gxa}
  J.~Louis and H.~Triendl,
  ``Maximally supersymmetric AdS$_{4}$ vacua in N = 4 supergravity,''
  JHEP {\bf 1410} (2014) 007
  [arXiv:1406.3363 [hep-th]].
  
\bibitem{Louis:2015mka}
  J.~Louis and S.~L\"ust,
  ``Supersymmetric AdS$_{7}$ backgrounds in half-maximal supergravity and marginal operators of (1, 0) SCFTs,''
  JHEP {\bf 1510} (2015) 120
  [arXiv:1506.08040 [hep-th]].
  

\bibitem{Louis:2015dca}
  J.~Louis, H.~Triendl and M.~Zagermann,
  ``$ \mathcal{N}=4 $ supersymmetric AdS$_{5}$ vacua and their moduli spaces,''
  JHEP {\bf 1510} (2015) 083
  [arXiv:1507.01623 [hep-th]].
  
  
\bibitem{Louis:2016qca}
  J.~Louis and C.~Muranaka,
  ``Moduli spaces of AdS$_{5}$ vacua in $ \mathcal{N} $ = 2 supergravity,''
  JHEP {\bf 1604} (2016) 178
  [arXiv:1601.00482 [hep-th]].

 
\bibitem{Louis:2016tnz}
  J.~Louis and S.~L\"ust,
  ``Classification of maximally supersymmetric backgrounds in supergravity theories,''
  arXiv:1607.08249 [hep-th].
  

\bibitem{Duff:1985jd}
  M.~J.~Duff and C.~N.~Pope,
  ``Consistent Truncations In Kaluza-klein Theories,''
  Nucl.\ Phys.\ B {\bf 255} (1985) 355.


\bibitem{Cassani:2010na}
  D.~Cassani and A.~F.~Faedo,
  ``A Supersymmetric consistent truncation for conifold solutions,''
  Nucl.\ Phys.\ B {\bf 843} (2011) 455
  [arXiv:1008.0883 [hep-th]].

\bibitem{Bena:2010pr}
  I.~Bena, G.~Giecold, M.~Grana, N.~Halmagyi and F.~Orsi,
  ``Supersymmetric Consistent Truncations of IIB on $T^{1,1}$,''
  JHEP {\bf 1104} (2011) 021
  [arXiv:1008.0983 [hep-th]].

\bibitem{Halmagyi:2011yd}
  N.~Halmagyi, J.~T.~Liu and P.~Szepietowski,
  ``On $N = 2$ Truncations of IIB on $T^{1,1}$,''
  JHEP {\bf 1207} (2012) 098
  [arXiv:1111.6567 [hep-th]].

\bibitem{Klebanov:1998hh}
  I.~R.~Klebanov and E.~Witten,
  ``Superconformal field theory on three-branes at a Calabi-Yau singularity,''
  Nucl.\ Phys.\ B {\bf 536} (1998) 199
  [hep-th/9807080].
  

\bibitem{Benvenuti:2005wi}
  S.~Benvenuti and A.~Hanany,
  ``Conformal manifolds for the conifold and other toric field theories,''
  JHEP {\bf 0508} (2005) 024
  [hep-th/0502043].
  
\bibitem{Ardehali:2014zfa}
  A.~Arabi Ardehali and L.~A.~Pando Zayas,
  ``On Exactly Marginal Deformations Dual to $B$-Field Moduli of IIB Theory on SE$_5$,''
  JHEP {\bf 1409} (2014) 164
  [arXiv:1405.5290 [hep-th]].
  

  


\bibitem{Lunin:2005jy}
  O.~Lunin and J.~M.~Maldacena,
  ``Deforming field theories with U(1) x U(1) global symmetry and their gravity duals,''
  JHEP {\bf 0505} (2005) 033
  [hep-th/0502086].

\bibitem{Hoxha:2000jf}
  P.~Hoxha, R.~R.~Martinez-Acosta and C.~N.~Pope,
  ``Kaluza-Klein consistency, Killing vectors, and Kahler spaces,''
  Class.\ Quant.\ Grav.\  {\bf 17} (2000) 4207
  [hep-th/0005172].

\bibitem{Amariti:2016mnz}
  A.~Amariti and C.~Toldo,
  ``Betti multiplets, flows across dimensions and c-extremization,''
  arXiv:1610.08858 [hep-th].
  
\bibitem{RHain}
	R.~Hain,
	``Lectures on moduli spaces of elliptic curves,''
	arXiv:0812.1803 [math.AG].  
  
\bibitem{Gunaydin:2000xk}
  M.~Gunaydin and M.~Zagermann,
  ``The Vacua of 5-D, N=2 gauged Yang-Mills/Einstein tensor supergravity: Abelian case,''
  Phys.\ Rev.\ D {\bf 62} (2000) 044028
  [hep-th/0002228].


\bibitem{Bergshoeff:2002qk}
  E.~Bergshoeff, S.~Cucu, T.~De Wit, J.~Gheerardyn, R.~Halbersma, S.~Vandoren and A.~Van Proeyen,
  ``Superconformal N=2, D = 5 matter with and without actions,''
  JHEP {\bf 0210} (2002) 045
  [hep-th/0205230].
  
\bibitem{Bergshoeff:2004kh}
  E.~Bergshoeff, S.~Cucu, T.~de Wit, J.~Gheerardyn, S.~Vandoren and A.~Van Proeyen,
  ``N = 2 supergravity in five-dimensions revisited,''
  Class.\ Quant.\ Grav.\  {\bf 21} (2004) 3015
   [Class.\ Quant.\ Grav.\  {\bf 23} (2006) 7149]
  [hep-th/0403045].

\bibitem{Andrianopoli:1996cm}
  L.~Andrianopoli, M.~Bertolini, A.~Ceresole, R.~D'Auria, S.~Ferrara, P.~Fre and T.~Magri,
  ``N=2 supergravity and N=2 superYang-Mills theory on general scalar manifolds: Symplectic covariance, gaugings and the momentum map,''
  J.\ Geom.\ Phys.\  {\bf 23} (1997) 111
  [hep-th/9605032].

\bibitem{Acharya:1998db}
  B.~S.~Acharya, J.~M.~Figueroa-O'Farrill, C.~M.~Hull and B.~J.~Spence,
  ``Branes at conical singularities and holography,''
  Adv.\ Theor.\ Math.\ Phys.\  {\bf 2} (1999) 1249
  [hep-th/9808014].

\bibitem{Cassani:2010uw}
  D.~Cassani, G.~Dall'Agata and A.~F.~Faedo,
  ``Type IIB supergravity on squashed Sasaki-Einstein manifolds,''
  JHEP {\bf 1005} (2010) 094
  [arXiv:1003.4283 [hep-th]].

\bibitem{Gauntlett:2010vu}
  J.~P.~Gauntlett and O.~Varela,
  ``Universal Kaluza-Klein reductions of type IIB to N=4 supergravity in five dimensions,''
  JHEP {\bf 1006} (2010) 081
  [arXiv:1003.5642 [hep-th]].

\bibitem{Liu:2011dw}
  J.~T.~Liu and P.~Szepietowski,
  ``Supersymmetry of consistent massive truncations of IIB supergravity,''
  Phys.\ Rev.\ D {\bf 85} (2012) 126010
  doi:10.1103/PhysRevD.85.126010
  [arXiv:1103.0029 [hep-th]].


\bibitem{Grimm:2014aha}
  T.~W.~Grimm, A.~Kapfer and S.~Lust,
  ``Partial Supergravity Breaking and the Effective Action of Consistent Truncations,''
  JHEP {\bf 1502} (2015) 093
  [arXiv:1409.0867 [hep-th]].
  
\bibitem{Sparks:2010sn}
  J.~Sparks,
  ``Sasaki-Einstein Manifolds,''
  Surveys Diff.\ Geom.\  {\bf 16} (2011) 265
  doi:10.4310/SDG.2011.v16.n1.a6
  [arXiv:1004.2461 [math.DG]].

\bibitem{Becker:2007zj}
  K.~Becker, M.~Becker and J.~H.~Schwarz,
  ``String theory and M-theory: A modern introduction,''

\bibitem{Klebanov:2000hb}
  I.~R.~Klebanov and M.~J.~Strassler,
  ``Supergravity and a confining gauge theory: Duality cascades and chi SB resolution of naked singularities,''
  JHEP {\bf 0008} (2000) 052
  [hep-th/0007191].
  

\bibitem{Gauntlett:2004yd}
  J.~P.~Gauntlett, D.~Martelli, J.~Sparks and D.~Waldram,
  ``Sasaki-Einstein metrics on $S^{2}\times S^{3}$,''
  Adv.\ Theor.\ Math.\ Phys.\  {\bf 8} (2004) no.4,  711
   [hep-th/0403002].

\end{thebibliography}
\providecommand{\href}[2]{#2}\begingroup\raggedright\endgroup

\end{document}